%Paper: hep-th/9504044
%From: Gregory Pelts <pelts@prufrock.Rockefeller.EDU>
%Date: Fri, 7 Apr 95 15:40:09 -0400
%Date (revised): Mon, 17 Apr 95 19:15:35 -0400
%Date (revised): Mon, 17 Apr 95 22:20:11 -0400
%Date (revised): Mon, 17 Apr 95 22:29:47 -0400

\input phyzzx
\PHYSREV
\hoffset=0.3in
\voffset=-1pt
\baselineskip = 14pt \lineskiplimit = 1pt
\frontpagetrue
\rightline {Cincinnati preprint October-1994}
\medskip
\titlestyle{\seventeenrm Quantum Stability of the Phase Transition
in Rigid QED}
\vskip.5in
\medskip
\centerline {\caps M. Awada \footnote*{\rm E-Mail
address:(moustafa@physunc.
phy.uc.edu}}
\centerline {\caps D.Zoller, and J.F.Clark}
\centerline {Physics Department}
\centerline {\it University of Cincinnati,Cincinnati, OH-45221}
\bigskip
\centerline {\bf Abstract}

Rigid QED is a renormalizable generalization of Feynman's space-time
action characterized by the addition of the curvature of the world
line (rigidity).  We have recently shown that a phase transition
occurs
in the leading approximation of the large N limit.  The disordered
phase
essentially coincides with ordinary QED, while the ordered phase is
a new theory. We have further shown that both phases of the quantum
theory are free of ghosts and tachyons. In this letter, we study the
first sub-leading quantum corrections leading to the renormalized
mass gap equation.  Our main result is that the phase transition does
indeed survive these quantum fluctuations.

\eject

{\bf I- Phase Transition in Rigid Model Of QED}

Recenlty we proposed a renormalizable generalization of the Feynman
space-time picture of QED [1], [2].  In this picture the dynamical
variables are the space-time position $x^{\mu}, \mu=1,2,...D$ of the
point particle and the photon field $A_{\mu}$.  The usual Feynman
action consists of the arc-length of the world line, the Maxwell
action, and the usual point particle-Maxwell coupling.
The renormalizable generalization is characterized by the addition
of the scale invariant curvature of the world line.  The origin of
the term rigid refers to the Boltzmann suppression of curved
trajectories by the curvature.

In a subsequent article [3] we proved a conjecture in [1] and [2]
that
there is phase transition in rigid QED.   We found a critical
line in the plane of the Coulomb coupling verses the curvature
coupling below which there is a disordered phase and above which is
a new ordered strongly coupled phase.
The higher derivative nature of rigid QED should cause a serious
pause
as any higher derivative theory is typically pathological. Indeed,
the arc-length plus the curvature term theory has classical runaway
solutions which are tachyonic.  Whether a higher derivative regulated
quantum theory has such  pathological behaviour is more subtle and
depends on details  of the continuum limit.  A free scalar field
theory on the lattice with spacing ${1\over \Lambda}$, has higher
derivatives and ghosts.  However these ghosts have mass of order
$\Lambda$ and decouple in the continuum limit as
$\Lambda\rightarrow\infty$.  In the less trivial case of rigid QED
we have shown that the ghosts have mass of order $\Lambda$ and
similarly
decouple from the continuum limit.  The necessity of the decoupling
mechanism is associated with the absence of fine tuning of the
curvature
and the Coulomb coupling constants.  Having phase transition would
be of utmost importance because this would imply that the couplings
of
the theory are fixed by dimensional transmutation in both the
disordered and ordered phases [4].

Our proof in [3] of the phase transition was based on the leading
order
approximation in large N, where N is the space-time dimensions.  Even
though,
the large N limit is a successful approximation
for non-linear sigma models, and some spin systems it can
sometimes lead to an incorrect conclusion.  The leading order
of the large N approximation is mean field theory which can give
incorrect predictions in lower dimensions.  For example mean field
theory incorrectly gives a phase transition in the one
dimensional Ising model. This discrepancy is
resolved by carefully examining the sub-leading quantum corrections
(loops) where one shows that such quantum corrections in fact destroy
the phase transition. Therefore it is crucial to examine
the quantum loop corrections to the mass gap, and the critical line
of
our model.

In this letter we will prove that the phase transition in our model
of rigid QED survives quantum fluctuations and that the quantum loop
corrections to the sub-leading order lead to mass and wave function
renormalizations.  As in non-linear sigma model [5], mass
renormalization
is equivalent to charge renormalization.  Thus we obtain the
renormalization group equation.

The effective action obtained after the Guassian integration of the
gauge field sector is [1], [2]:
$$ \mu_{0}\int_{0}^{1} d\lambda \sqrt{{\dot x}^{2}}+
{1\over t_{0}}\int_{0}^{1} d\lambda {\sqrt{{\dot x}^{2}
{\ddot x}^{2}-({\dot x}.{\ddot x})^{2}}
\over {\dot x}^{2}}+ {1\over 2t_{0}}\int_{0}^{1}\int_{0}^{1}
d\lambda d\lambda' {\dot x}(\lambda){\dot x}(\lambda')
V(|x-x'|)\eqno{(1 a)}$$
where the first term is the arc-length $ds=
d\lambda \sqrt{{\dot x}^{2}}$ of a point particle of bare
mass $\mu_{0}$, the second term is the curvature
k(s) = $| {d^{2}x(s)\over ds^{2}} |$  of the world line defined to
be the length of the acceleration, $t_{0}$ is  a dimensionless bare
coupling constant (scale invariance of the curvature term) and V is
the long range Coulomb potential:
$$ V(|x-x'|,a)= {2g_{0}\over \pi}
{1\over |x(\lambda)-x(\lambda')|^{2}+a^{2}}\   .\eqno{(1 b)}$$
We have introduced the cut-off "a" to avoid the singularity at
$\lambda=\lambda'$ and define $ g_{0}=t_{0}.\alpha_{Coulomb}=
t_{0}.{e_{0}^{2}\over 4\pi}$.  In the arc-length gauge ${\dot
x}^{2}=1$
we can gauge fix the action (1a) and obtain:
$${\tilde S} ={1\over 2}\mu_{0}L +{1\over 2t_{0}}\int_{0}^{L}
d\lambda(e^{-1}{\ddot x}^{2} + e + \omega({\dot x}^{2} - 1) )
         + {1\over 2t_{0}}\int_{0}^{L}\int_{0}^{L} d\lambda d\lambda'
           {\dot x}(\lambda){\dot x}(\lambda')V(|x-x'|,a)\eqno{(2)}$$
where e is an einbein to remove the square root of the acceleration,
$\omega$ is a Lagrange multiplier that enforces the constraint
${\dot x}^{2}=1$, and L is the length of the path.  The partition
function is:
$$ Z =\int D\omega De Dx exp(-{\tilde S})\  .\eqno{(3)}$$
The long range Coulomb interactions are non-local and
impossible to integrate.  Therefore we consider
$$x^{\nu}(\lambda) = x^{\nu}_{0}(\lambda) + x^{\nu}_{1}(\lambda)$$
and expand the action (2) to quadratic order in
$x^{\mu}_{1}(\lambda),\mu=1,...D$
about the background straight line $x_{0}$.  The x-integration is now
Gaussian and to the leading D approximation we obtain the following
full effective action $S_{eff}$:
$$ {S}_{eff} = {1\over 2t_{0}}\int d\lambda
e(\lambda)-\omega(\lambda) + t_{0}Dtrln A\eqno{(4)}$$
where A is the operator
$$ A = \partial^{2}e^{-1}\partial^{2} -\partial\omega\partial
+ V(\lambda,\lambda')\  .\eqno{(5)}$$
In the large D limit the stationary point equations resulting from
varying $\omega$ and e respectively are:
$$1=t_{0}DtrG\eqno {(6 a)}$$
$$1=t_{0}Dtr(e^{-2}(-\partial^{2}G))\eqno {(6 b)}$$
where the world line Green's function is defined by:
$$ G(\lambda,\lambda') = <\lambda|(-\partial^{2})A^{-1}|\lambda'>
\eqno{(7)}$$
The stationary points are:
$$ \omega(\lambda) = \omega_{0},~~~~~~<\lambda|e^{-1}|\lambda'>=
\int{dp\over 2\pi} {e^{i(\lambda-\lambda')}\over |p|}\eqno{(8)}$$
where $\omega_{0}$ is the bare mass which is a positive constant.
Using the stationary solutions (8) the operator A is given by:
$$ trln A = \int{dp\over 2\pi}ln[|p|^{3} + p^{2}\omega_{0} +
p^{2}V_{0}(|p|) + V_{1}(|p|)]\eqno{(9)}$$
where
$$ V_{0}(|p|) = {2g_{0}\over a}e^{-a|p|},~~~~~V_{1}(|p|)={2g_{0}
\over a^{2}}[e^{-a|p|}(|p|+ {1\over a}) -{1\over a}]\  .\eqno{(10)}$$
Thus Eq.(6) becomes the single mass gap equation [3]:
$$ 1 =Dt_{0}\int{dp\over 2\pi}{p^{2}\over p^{2}(|p|+\omega_{0}) +
p^{2}V_{0}(|p|) + V_{1}(|p|)}\eqno{(11)}$$
%%%$$ \omega^{*} = \Lambda e^{-{\pi\over Dt}}\eqno{(10)}$$
where $\omega_{0}$ is now the mass associated with the propagator:
$$ <{\dot x}^{\mu}(p){\dot x}^{\nu}(-p)>
 = Dt_{0}{\delta^{{\mu}{\nu}}\over |p|+\omega_{0} + {\cal V}(|p|)}
\eqno{(12)}$$
where:
$${\cal V}(|p|) = V_{0}(|p|) + {V_{1}(|p|)\over p^{2}}\eqno{(13)}$$
which is regular at p =0.
To obtain a  non-zero phase transition temperature the mass gap
equation must be infra-red finite for $\omega_{0}=0$.  Therefore
without Coulomb long range interactions ($g_{0}=0$) the theory exists
only in the disordered phase $t_{0}>t_{c}$ and the U.V stable fixed
point is $t_{c}=0$.  In this case the beta function of the
pure curvature theory is asymptotically free, indicating the
absence of the curvature term at large distance scales.  In contrast
to the naive classical
limit the theory is therefore well behaved and free of ghosts.
In  [3] we calculated the poles of the Green's
function in the presence of Coulomb interactions ($g_{0}\neq 0$)
using the large D limit, and showed the absence of ghosts and
tachyons
in both the ordered and disordered phases which are separated by the
critical line defined by eq.(11) at $\omega_{0}=0$:
$$ 1 ={Dt_{0}\over \pi}\int_{0}^{\rho} dy{y^{2}\over y^{3}
+2g_{0}(y^{2}e^{-y}+ye^{-y}+e^{-y}-1)}\eqno{(14)}$$
where we have made the change of variable y=ap and introduced the
U.V.
cut-off $\Lambda$, and $\rho = {\Lambda\over \Lambda_{0}}$ where
${1\over a}:=\Lambda_{0}$.  It is straightforward to prove that there
exist
a $g_{0}^{*}$ ( c.f. Fig.1) for which any choice of $\rho$
leads to phase transition as long as $g_{0}<g_{0}^{*}$.  We will set
$\rho =1$.  Notice that eq.(14) is finite except at $g_{0}=0$
(absence of Coulomb interactions).
The critical curve distinguishing the two phases in the (t,g) plane
is shown in Fig.1.  The order parameter of the theory is the mass gap
equation (14) where $\omega_{0}$ is the parameter that
distinguishes the two phases.  In the disordered phase
$\omega_{0} >0$, while in the ordered phase it is straightforward
to show that $\omega_{0} =0$.   In the disordered phase the coupling
constants $t_{0}$ and $g_{0}$ are completely fixed by dimensional
transmutation in terms of the cut-off $\Lambda$ and $\omega_{0}$.
Thus, they cannot be fine tuned.  This is an important property that
is vital in proving the absence of ghosts in our model [3].  From
(14)
we can immediately examine whether there is a phase transition in
the pure Coulomb theory i.e ordinary QED.  The curvature term would
then be absent.  This corresponds to the absence of the $y^3$ term
in (14).  If we choose the cut-off of the theory $\Lambda$ to be at
the Compton wavelength i.e $\Lambda_{Compton}={2\pi\over a}$ one
finds in this particular case that the integral (14) diverges
implying an absence of a phase transition.
\eject

{\bf II- The Loop Corrected Gap Equation}

In mean field theory i.e leading order in ${1\over D}$,
the relevant propagator is equation (12).  In the sub-leading
correction to mean field theory, the quantum fluctuation
imply a new term corresponding to the self-energy of the
${\dot x}^{\mu}$-field
$$ <{\dot x}^{\mu}(p){\dot x}^{\nu}(-p)>
 = Dt_{0}{\delta^{{\mu}{\nu}}\over (|p|+\omega_{0} + {\cal V}(|p|)
+ {1\over D}\Sigma(|p|))}\  .\eqno{(15)}$$
The new contribution $\Sigma (|p|)$ arises from fluctuations of the
Lagrange multiplier $\omega$ where the fluctuations $\eta$ are
defined by:
$$ \omega = \omega_{0} + i{1\over \sqrt{(D/2)}} \eta\  .\eqno{(16)}$$
Expanding the effective action in powers of $\eta$, it is
straightforward
to extract the $\eta$ propagator (Fig (2)):
$$ \Pi(|p|) = \int_{-\infty}^{+\infty} {dk\over 2\pi} {1\over (|k| +
\omega_{0} +  {\cal V}(|k|))}{1\over (|p + k| +
\omega_{0} +  {\cal V}(|p + k|))}\  .\eqno{(17)}$$
The self energy $\Sigma$ can be computed from the diagrams of
Fig(3).  These diagrams are of order ${1\over D}$ and represent
the quantum fluctuations:
$$\Sigma(|p|) = \int_{-\infty}^{+\infty} {dk\over 2\pi}
{\Pi^{-1}(|k|)\over (|p + k| +
\omega_{0} +  {\cal V}(|p + k|))}$$
$$ - \int_{-\infty}^{+\infty}
{dk\over 2\pi} \int_{-\infty}^{+\infty} {dq\over 2\pi} {\Pi^{-1}(0)
\over (|q| + \omega_{0} + {\cal V}(|q|))^{2}}
{\Pi^{-1}(|k|)\over (|q + k| +
\omega_{0} + {\cal V}(|q + k|))}\  .\eqno{(18)}$$
A Taylor expansion of the self energy about zero momentum leads
to mass and wave function renormalizations and a remaining piece
${\tilde \Sigma}$. The propagator now reads:
$${Z\over (|p|+\omega + {\cal V}(|p|)
+ {1\over D}{\tilde \Sigma(|p|)})}\  .\eqno{(19)}$$
where
$$ Z = 1 - {1\over D} \Sigma'(0)\eqno{(20 a )}$$
is the wave function renormalization,
$$\omega =\omega_{0} + {1\over D}(\Sigma(0) -\omega_{0}
\Sigma'(0))\eqno{(20 b )}$$
is the renormalized mass and,
$$ g = Z g_{0}\eqno{(20 c)}$$
is the renormalized Coulomb coupling constant.  The renormalized
mass gap equation is
$$ 1 = Dt \int_{-\infty}^{+\infty}{dp\over 2\pi}{1\over (|p|+\omega +
{\cal V}(|p|) + {1\over D}{\tilde \Sigma(|p|)})}\  .\eqno{(21)}$$
To examine whether there is still a phase transition
i.e the infra-red finiteness of the renormalized mass gap equation
eq.(21) at $\omega = 0$, we must examine the renormalizability of the
theory.   We have seen that the typically divergent terms $\Sigma(0)$
and $\Sigma'(0)$ can be absorbed in mass and wave function
renormalizations.
This will be true if we can regularize (18) so as to respect
(19) and (20) with ${\tilde \Sigma(|p|)}$ being finite.
Equivalently,
$\Sigma''(|p|)$ must be ultra-violet (U.V.) finite.  To study this
question we only need the asymptotic behaviour of $\Pi$ :
$$ \Pi_{as}(|p|) = {2\over \pi}{log{|p|\over \omega_{0}}\over ((|p| +
{\cal V}(|p|))}\  .\eqno{(22)}$$
{}From (18) it is straightforward to find that
$$\Sigma''(|p|) = 2 \int_{-\infty}^{+\infty}
{dk\over 2\pi}(1 + {\cal V}'(|p+k|))^{2}
{\Pi^{-1}(|k|)\over (|p + k| + \omega_{0} + {\cal V}(|p + k|))^3}$$
$$-{\Pi^{-1}(|p|)\over \pi}{(1-{4\over 3 }g_{0})\over
(\omega_{0} + {g_{0}\over a})^{2}} -\int_{-\infty}^{+\infty}
{dk\over 2\pi}{\cal V}''(|p+k|)
{\Pi^{-1}(|k|)\over (|p + k| + \omega_{0} + {\cal V}(|p + k|))^2}
\  .\eqno{(23)}$$
For a non zero cut-off "a" it is clear from (23) that the second term
is a finite function since $\Pi^{-1}(|p|)$ is a finite function of
$|p|$.
Inserting (22) into (23) one can easily prove that $\Sigma''(0)$
is ultra-violet (u.v) finite and thus $\Sigma''(|p|)$ is a finite
function of $|p|$.  Furthermore, although ${\cal V}(a,|p|)$
diverges as ${-2g_{0}\over a^3 p^{2}}$ as $a \rightarrow 0$, one can
check that $\Sigma''(|p|)$ is still U.V. finite in this limit.  Due
to the
complexity of the long range potential ${\cal V}(a,|p|)$ we cannot
calculate $\Sigma''(|p|)$ exactly.  Therefore, to further
confirm whether $\Sigma''(|p|)$ is indeed U.V. finite, Eq.(23) is
calculated  numerically for arbitrary fixed values of the cut-off a
and
mass $\omega$. Fig.4 depicts a three-dimensional-surface calculation
of
$\Sigma''(|p|)$ in the p-g plane for a particular choice of a and
$\omega$.
To facilitate numerical calculations, $\Sigma''(|p|)$ is plotted in
the
range $0<g<g^*$.  In this range there are no real poles in
the propagator (12) since  $(|p|+{\cal V}(|p|)) > 0$ for arbitrary p.
Different choices of a and $\omega$
reveal similar smooth surfaces, and, as $g\rightarrow g^*$
$\Sigma''(|p|)$
becomes more negative but always remains finite for arbitrary p. Due
to the
asymptotic form of $\Pi_{as}(|k|)$ defined in Eq.(22), care must be
taken in
the numerical integration over k in (23) to avoid artificial
singularities in
the small k regime. Alternatively we can work out the regularized and
finite
self energy using the SM regularization scheme [5]:
$$ {\tilde \Sigma}_{finite}(|p|) = \Sigma(|p|) -{\omega_{0}\over
2}(I_{0}
-{1\over \omega_{0}}I_{1}) +{1\over 2}(|p| +
\omega_{0})I_{0}\eqno{(24 a)}$$
$$ I_{0}({\Lambda\over \omega_{0}},g)=
\int_{{\tilde \Lambda}}^{\Lambda} dk {1\over log{k\over \omega_{0}}}
{1\over (|k|+{\cal V}(|k|))}\eqno{(24 b)}$$
$$ I_{1}({\Lambda\over \omega_{0}},g) =
\Pi(0)^{-1}\int_{{\tilde \Lambda}}^{\Lambda}{dk\over \pi} {1\over
(|k|+
\omega_{0} + {\cal V}(|k|))}={2\Pi(0)^{-1}\over Dt_{0}}\  .\eqno{(24
c)}$$
To the first sub-leading order we have therefore shown that the
theory is
renormalizable. The renormalization group flow for the renormalized
curvature coupling $t$ and the Coulomb coupling $g$ can be derived
from
(20c) and the mass renormalization by holding $\omega(\Lambda,t,g)$
fixed.  From (11), (18) and (20) and (24) we obtain:
$$\omega = {\tilde Z}\omega_{0} = (Z - {I_{1}({\Lambda\over
\omega_{0}},g)
\over D\omega_{0}})\omega_{0}\eqno{(25 a)}$$
where Z is given by:
$$Z = 1 +{1\over 2D} I_{0}({\Lambda\over \omega_{0}},g)\  .\eqno{(25
b)}$$
Having shown that ${\tilde \Sigma}$ is indeed finite then the loop
corrected critical line of the theory defined by (21) at $\omega = 0$
is:
$$ 1 = Dt \int_{-\infty}^{+\infty}{dp\over 2\pi}{1\over (|p|+
{\cal V}(|p|) + {1\over D}{\tilde \Sigma_{finite}(|p|)})}\
.\eqno{(26)}$$
which is Infra-red finite since by its definition ${\tilde
\Sigma(0)}=0$
and (26) then has the exact behaviour as (14) at y=0.  Furthermore,
as long
as $g<g^{*}$ we have shown that the critical line (14) has no real
poles.
i.e $(|p|+{\cal V}(|p|)) > 0$. The presence of ${\tilde \Sigma}$ will
not
affect the above conclusion to any sub-leading finite order in
perturbation
theory. In the sub-leading ${1\over D}$ order the critical line is:
$$ 1 = Dt \int_{-\infty}^{+\infty}{dp\over 2\pi}({1\over (|p|+
{\cal V}(|p|))} - {1\over (|p|+{\cal V}(|p|))}{1\over D}{\tilde
\Sigma_{finite}(|p|)}{1\over (|p|+{\cal V}(|p|))})\  .\eqno{(27)}$$
As in any quantum field theory the location of the poles in the
presence
of ${\tilde \Sigma}_{finite}(|p|)$ is a non-perturbative issue and
requires the form of ${\tilde \Sigma}_{finite}(|p|)$ to all orders in
perturbation theory.

{\bf Acknowledgement}

 We are very grateful to Prof. A. Polyakov for his constant
encouragement and extensive support over the last year and a half and
for
suggesting that we address the quantum stability of the phase
transition both
in the model of rigid QED and that of rigid strings coupled to long
range
interactions [6]. In addition, one of us M.A is thankful for his long
and
patient conversations explaining the decoupling mechanism of ghosts
and its
connection with critical phenomena. We are also grateful to Prof. Y.
Nambu
for his extensive support, encouragement and long discussions over
the last
two years without which we could not have gone far in our
investigations.
M.A would like to thank P.Ramond, C. Thorn, and Z. Qui for
constructive
discussions and suggestions.

{\bf References}

\item {[1]} M. Awada and D. Zoller, Phys.Lett B299 (1993) 151,
 M.Awada, M.Ma, and D.Zoller Mod.phys. Lett A8,(1993), 2585
\item {[2]} M.Awada and D.Zoller, Int. J. Mod. Phys. A9(1994)
pp.4077-4099.
\item {[3]} M. Awada and D. Zoller, Phys.Lett B325 (1994) 119
\item {[4]} A. Polyakov, private communications.
\item {[5]} A. Polyakov,  Gauge fields, and Strings,
Vol.3, harwood academic publishers, J.Orloff and R.Brout, Nucl.
Phys. B270 [FS16],273 (1986), M. Campostrini and P.Rossi, Phys.
Rev.D 45, 618 (1992) ; 46, 2741 (1992), H. Flyvberg, Nucl. Phys.
B 348, 714, (1991).
\item {[6]} M. Awada and D. Zoller, Phys.Lett B325 (1994) 115

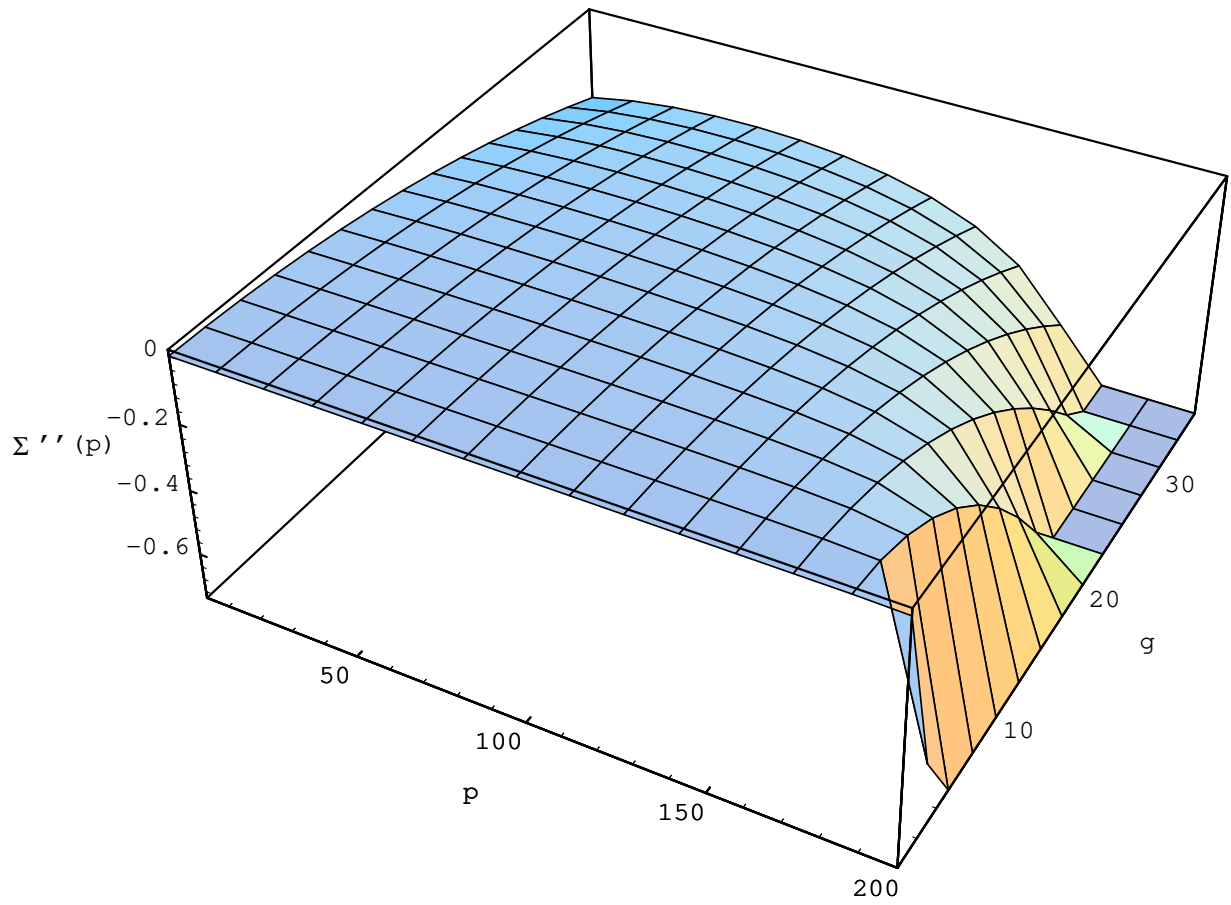
\end